\documentclass{sf2a-conf}
\usepackage{graphicx}
\newcommand{\eg} {{\em e.g.}}
\newcommand{\HH}   {\mbox{H$_2$}}       
\newcommand{\twCO} {\mbox{$^{12}$CO}}   
\newcommand{\CeiO} {\mbox{C$^{18}$O}}   
\newcommand{\DCOp} {\mbox{DCO$^{+}$}} 
\newcommand{\HthCOp} {\mbox{H$^{13}$CO$^{+}$}} 
\newcommand{\CCH}{\mbox{CCH}}             %
\newcommand{\CCCCH}{\mbox{C$_4$H}}        %
\newcommand{\cCCCHH}{\mbox{c-C$_3$H$_2$}} %
\newcommand{\ion}[2]{\mbox{#1{\sc #2}}}
\newcommand{\CI} {\ion{C}{i}}       
\newcommand{\Jone}{\mbox{J=1--0}}
\newcommand{\Jtwo}{\mbox{J=2--1}}
\newcommand{\Jthr}{\mbox{J=3--2}}
\newcommand{\Jfou}{\mbox{J=4--3}}

\newcommand{\emm}[1]{\ensuremath{#1}}   
\newcommand{\emr}[1]{\emm{\mathrm{#1}}} 
\newcommand{\unit}[1]{\emm{\, \emr{#1}}}
\newcommand{\Myr} {\unit{Myr}}
\newcommand{\K}   {\unit{K}}
\newcommand{\Kpccm}{\unit{K\,cm^{-3}}}
\newcommand{\mm}  {\unit{mm}}

\newcommand{\mim} {\unit{\mu m}}
\newcommand{\pccm}{\unit{cm^{-3}}}

\newcommand{\kms}{\unit{km\,s^{-1}}}
\newcommand{\kmsppc}{\unit{km\,s^{-1}\,pc^{-1}}}

\renewcommand{\deg}{\emm{^\circ}}

\newcommand{\pc}    {\unit{pc}}

\newcommand{\Msun}  {\unit{M_\odot}}

\renewcommand{\deg}{\emm{^\circ}}

\newcommand{\about}{\emm{\sim}} 
\begin{document}

\TitreGlobal{SF2A 2006}

\title{Benchmarking PDR models against the Horsehead edge}
\author{Pety, J.}\address{IRAM, 300 rue de la Piscine, 38406 Grenoble
  cedex, France}
\author{Goicoechea, J.R.}\address{LERMA-LRA, UMR 8112, CNRS, Observatoire
  de Paris and ENS, 24 rue Lhomond, 75231 Paris cedex 05, France}
\author{Gerin, M.$^2$}
\author{Hily-Blant, P.$^1$}
\author{Teyssier, D.}\address{ESAC, Urb. Villafranca del Castillo, P.O. Box
  50727, Madrid 28080, Spain}
\author{Roueff, E.}\address{LUTH, UMR 8102, CNRS and Observatoire de Paris,
  Place J. Jansen, 92195 Meudon cedex, France}
\author{Habart, E.}\address{IAS, Universit\'e Paris-Sud, B\^at. 121, 91405
  Orsay, France}
\author{Abergel, A.$^5$}
\runningtitle{Benchmarking PDR models against the Horsehead edge}
\setcounter{page}{237}
\index{Pety, J.}
\index{Goicoechea, J.}
\index{Gerin, M.}
\index{Hily-Blant, P.}
\index{Teyssier, D.}
\index{Roueff, E.}
\index{Habart, E.}
\index{Abergel, A.}

\maketitle
\begin{abstract}
  To prepare for the unprecedented spatial and spectral resolution provided
  by ALMA and Herschel/HIFI, chemical models are being benchmarked against
  each other. It is obvious that chemical models also need
  well-constrained observations that can serve as references.
  Photo-dissociation regions (PDRs) are particularly well suited to serve
  as references because they make the link between diffuse and molecular
  clouds, thus enabling astronomers to probe a large variety of physical
  and chemical processes.  At a distance of 400 pc ($1''$ corresponding to
  0.002\pc{}), the Horsehead PDR is very close to the prototypical kind of
  source (\ie{} 1D, edge-on) needed to serve as a reference to models.
\end{abstract}
%
\vspace*{-0.20cm}
\section{Introduction}
\vspace*{-0.25cm}

Photodissociation region models are used to understand the evolution of the
far UV illuminated interstellar matter both in our Galaxy and in external
galaxies. To prepare for the unprecedented spatial and spectral
capabilities of ALMA and Herschel, two different kinds of advances are
currently taking place in the field.  First, numerical models describing
the chemistry of a molecular cloud are being benchmarked against each
others to ensure that all models agree not only qualitatively but also
quantitatively on at least simple cases
(\texttt{http://www.ph1.uni-koeln.de/pdr-comparison}). Second, new or
improved chemical rates are being calculated/measured by several
theoretical and experimental groups. However, the difficulty of this last
effort implies that only a few reactions (among the thousands used in
chemical networks) can be thoroughly studied. This led to several recent
studies trying to identify a few key chemical reactions in some
well-studied astrophysical cases by comparison of ``observed'' and modelled
abundances, taking into account both the observational and chemical rate
uncertainties (\eg{} Wakelam et al. 2004, 2006). In view of the intrinsic
complexity of building reliable chemical networks and models, there is an
obvious need of well-constrained observations that can serve as basic
references. PDRs are particularly well suited to serve as references
because they make the link between diffuse and dark clouds, thus enabling
to probe a large variety of physical and chemical processes.

The Horsehead nebula, also known as Barnard 33, is one of the most famous
object of the sky. For instance, it was selected by internet voters as a
target for the Hubble Space Telescope in the framework of its 11th
anniversary (See the Hubble Heritage site for details). But it is also a
fantastic physical and chemical laboratory which has been actively studied
in the past 5 years.

\vspace*{-0.20cm}
\section{The Horsehead nebula as a typical pillar in star forming regions}
\vspace*{-0.25cm}

\begin{figure}[t]
  \centering
   \includegraphics[height=0.8\textwidth{},angle=270]{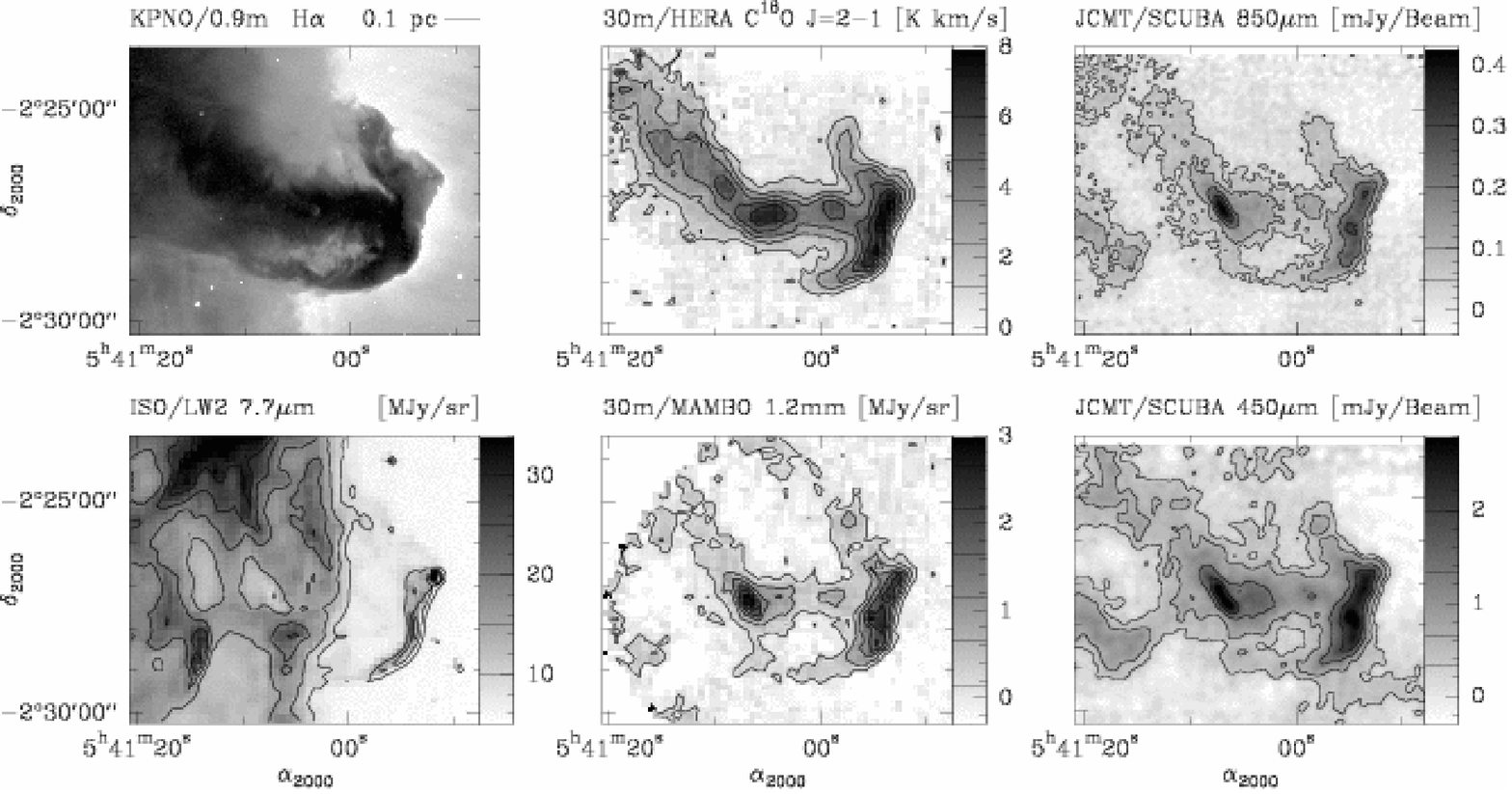}
  \caption{$500'' \times 400''$ maps of the Horsehead nebula in different
    tracers: H$\alpha$ (resolution: $\sim1''$, courtesy of Reipurth \&
    Bally), 7.7\mim{} aromatic continuum (resolution: $6''$, Abergel et al.
    2003), 9.1--11.8\kms{} \CeiO{}~\Jtwo{} emission and 1.2\,mm dust
    continuum (resolution: $11.7''$, Hily-Blant et al. 2005), 850\mim{} and
    450\mim{} dust continuum (resolution: $15''$, courtesy of G.~Sandell,
    published in Ward-Thompson et al. 2006). Values of contour level are
    shown on each image lookup table.}
  \label{fig:environment}
\end{figure}

The Horsehead nebula appears on optical images as a dark patch of $\sim 5'$
extent against the bright HII region IC~434. Mid-infrared to submillimeter
wavelength emissions from the gas and dust associated with this globule
clearly indicate its morphology (see figure~1). A curved rim including the
nose, ridge and mane of the horse is located at the western end of the
nebula (\ie{} the top of the head). This ridge is connected to the eastern
L1630 molecular cloud by a thin filament, the horse throat. While optically
thick tracers like the \twCO{} lines show a low density halo surrounding
the throat traced by the \CeiO{}~\Jtwo{} emission, optically thin tracers
like the dust emission (from 1.2\mm{} to 350\mim{}) reveal the presence of
two dense condensations: the first one associated with the western ridge,
called B33-SMM1, and the second one in the middle of the throat, called
B33-SMM2 (the \twCO{} and 350\mim{} images may be found in Pound et al.
2003 and in Philipp et al. 2006).

Two massive stars are located at the same projected distance of the
Horsehead (0.5\deg{} or 3.5\pc{} at a distance of 400\pc{}), namely the
O9.6Ib star $\zeta$Ori north of the nebula and the O9.5V star $\sigma$Ori
west of the nebula. However, figure~1 of Pound et al. (2003) shows that
some molecular gas is located in projection between $\zeta$Ori and the
Horsehead nebula. Moreover, Hipparcos estimates of their distance (Perryman
et al. 1997) indicate that $\zeta$Ori ($250\pm50\pc$) is located further
away from the Horsehead nebula than $\sigma$Ori ($352\pm113\pc$). Finally,
Philipp et al. (2006) present in their figure~6 the image of the
\twCO{}~\Jfou{}/\Jtwo{} ratio, which is an excellent tracer of the
direction of the incoming far UV photons: Indeed, this ratio increases at
the PDR edges due to the difference of optical depth between the two
\twCO{} transitions in regions with a steep temperature gradient.  The
\twCO{}~\Jfou{}/\Jtwo{} image clearly shows that the far UV illumination
primarily comes from $\sigma$Ori, even though $\zeta$Ori probably plays a
role in the north rim of the nebula (\ie{} the top of the neck). 

Pound et al. (2003) were the first to cast the Horsehead nebula in the
category of the pillars like the well-known examples in the Eagle nebula.
They guess that the Horsehead nebula formed through the photoevaporation of
low density material around the neck which was protected by the shadow of
denser material in the ridge. The typical size and velocity gradients thus
imply a formation timescale of \about{} 0.5\Myr{} and a timescale of
destruction through photoablation of \about{} 5\Myr{}. Hily-Blant et al.
(2005) showed through careful measurement of velocity gradients in the
\CeiO{}~\Jtwo{} emission that the gas is rotating around the neck axis. The
velocity gradients perpendicular to the neck axis are reasonably constant
around an average value of 1.5\kmsppc{} (implying a rotation period of
4\Myr{}) except at the position of the dense condensation B33-SMM2, where
the transverse velocity gradients experience a sharp increase up to
4\kmsppc{}.  The overall shape of the throat is thus assumed to be
cylindrical with a fairly constant projected diameter of about
0.15--0.30\pc{}. Since this study, Gahm et al. 2006 have confirmed that
rotation of gas around pillar major axes is a common phenomenon.
Hily-Blant et al. (2005) argued that the famous shape of the Horsehead
nebula could derive from a pre-existent velocity field that progressively
separated the mane and nose from the neck via the centrifugal effect.

\begin{figure}[t]
  \centering
  \includegraphics[width=0.8\textwidth{}]{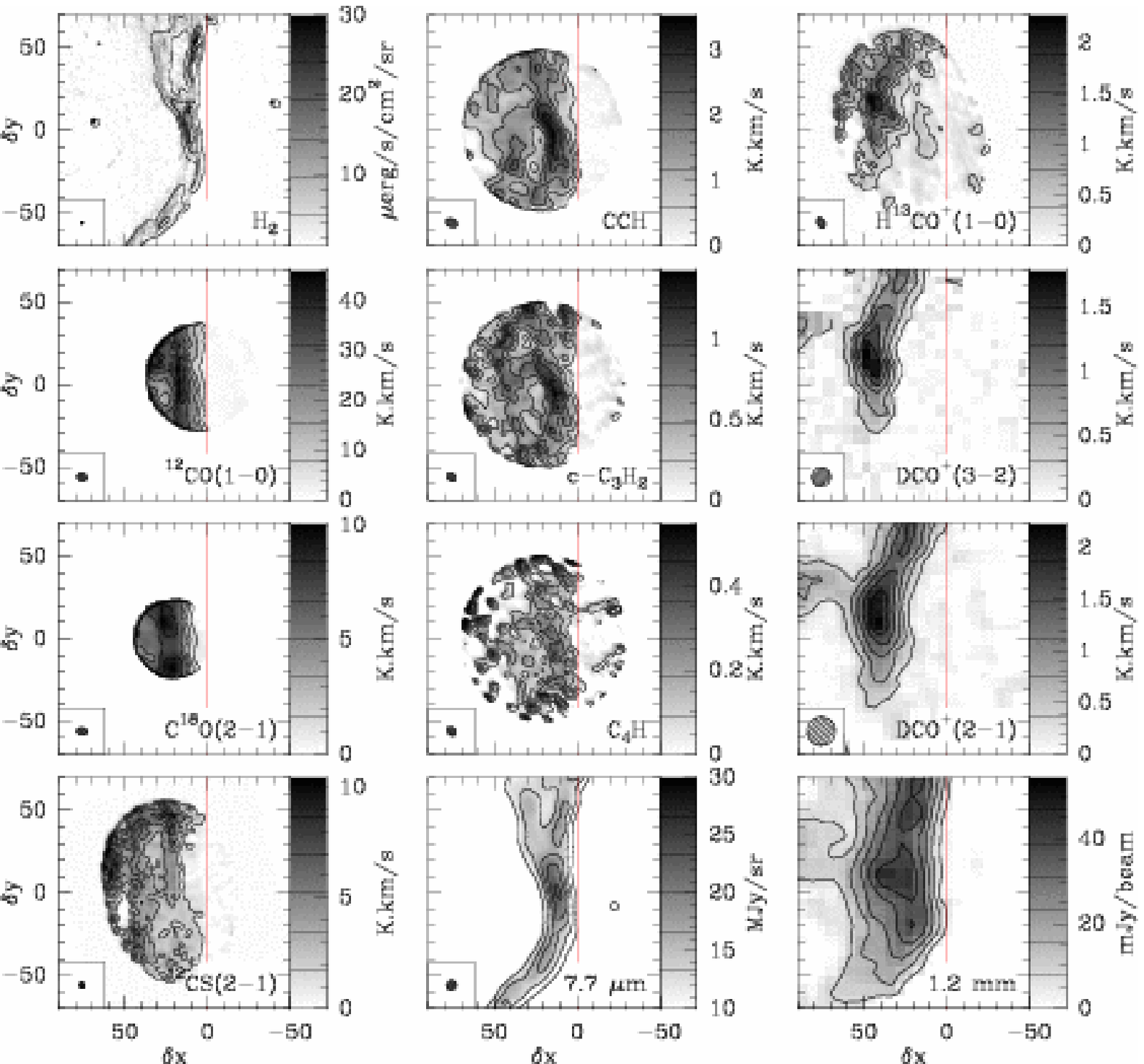}
  \caption{Emission maps obtained with the IRAM Plateau de Bure 
    Interferometer or 30m single-dish, except for the \HH{} v=1-0 S(1)
    emission observed with the NTT/SOFI and the mid--IR emission observed
    with ISO-LW2. The image center is RA(2000) = 05h40m54.27s, Dec(2000) =
    -02\deg{}28'00$''$. The maps have been rotated by 14\deg{}
    counter--clockwise around the image center to bring the exciting star
    direction in the horizontal direction as this eases the comparison of
    the PDR tracer stratifications. Maps have also been horizontally
    shifted by $20''$ to set the horizontal zero at the PDR edge delineated
    as the vertical red line. Either the synthesized beam or the single
    dish beam is plotted in the bottom left corner. The emission of all the
    lines is integrated between 10.1 and 11.1~\kms{}. Values of contour
    level are shown on each image lookup table (contours of the \HH{} image
    have been computed on an image smoothed to $5''$ resolution).}
  \label{fig:pdr}
\end{figure}

In their simultaneous study of the CO and \CI{} emissions, Philipp et al.
(2006) deduce a total molecular mass of 24\Msun{} in the throat and
13\Msun{} in the outer halo. This leads to an upper limit (as the regions
traced by \CeiO{} and \CI{} may partially overlap) of the total mass of
\about{} 37\Msun{}. Ward-Thompson et al. (2006) derived the main
characteristics of the two condensations of the Horsehead nebula. The
eastern condensation (B33-SMM2) is seen in emission in the (sub)millimeter
range and in absorption in the mid-infrared domain. Assuming a temperature
of 15\K{}, its mass is \about{} 4\Msun{} in a region of dimension $0.15
\times 0.07\pc$. This yields an average \HH{} density of $10^5\pccm$ while
the peak density is $2\times 10^6\pccm$. Virial estimates imply that this
dense core is in approximate gravitational equilibrium. The size of the
western condensation (B33-SMM1) is $0.13 \times 0.31\pc$ and its mass is
\about{} 2\Msun{} (assuming a temperature of 30\K{}). This yields an
average \HH{} density of $10^4\pccm$ while the peak density is $6\times
10^5\pccm$. Its virial balance is however largely dominated by the ionizing
radiation. Both condensations are potential progenitors of the next
generation of star formation and may thus give clues about the differences
between triggered and spontaneous star formation.

\vspace*{-0.20cm}
\section{The Horsehead PDR as a chemical laboratory}
\vspace*{-0.25cm}

The illuminated edge (PDR) of the western condensation presents one of the
sharpest infrared filament (width: $10''$ or 0.02\pc{}) detected in our
Galaxy by ISOCAM. The most straightforward explanation given by Abergel et
al. (2003) is that most of the dense material is within a flat structure
viewed edge-on and illuminated in the plane of the sky by $\sigma$Ori.  The
\HH{} fluorescent emission observed by Habart et al. (2005) is even sharper
(width: $5''$), implying the inclination of the PDR on the plane-of-sky to
be less than 5\deg{}. The Horsehead ridge thus offers the opportunity to
study at small linear scales ($1''$ corresponds to 0.002\pc{} at 400\pc{})
the physics and chemistry of a PDR with a simple geometry, very close to
the prototypical kind of source needed to serve as a reference to chemical
models. Since 2001, we started to study the Horsehead PDR mainly with the
IRAM Plateau de Bure interferometer at 3\mm{} and the IRAM-30m at 1\mm{}
achieving spatial resolutions from 3 to $11''$ (except the \HH{}
fluorescent emission observed at $1''$-resolution with the NTT/SOFI
instrument). Figure~2 displays all the observed, high resolution maps
already acquired. Those maps trace the different layered structures
predicted by photochemical models according to chemistry networks,
excitation conditions and radiative transfer.

Abergel et al. (2003) deduced from the distance between $\sigma$Ori and the
PDR that the intensity of the incident far UV radiation field is $\chi \sim
60$ relative to the interstellar radiation field in Draine's units.
Through the modelling of the \HH{} and CO emission, Harbart et al. (2005)
showed that the PDR has a very steep density gradient, rising to
$n_\emr{H} \sim 10^5\pccm$ in less than $10''$ (\ie{} 0.02\pc{}), at a
roughly constant pressure of $P \sim 4 \times 10^6\Kpccm$. These
observations were followed by a chemical study of small hydrocarbons
(\CCH{}, \cCCCHH{}, \CCCCH{}).  Pety et al. (2005) showed that the
abundances of the hydrocarbons are higher than the predictions based on
pure gas phase chemical models (Le Petit et al. 2002 and reference
therein).  These results suggest that, in addition to gas-phase chemistry,
another formation path of carbon chains should be considered in PDRs.  Due
to the intense far UV illumination, large aromatic molecules and small
carbon grains may fragment and feed the interstellar medium with small
carbon clusters and molecules in significant amounts. Recently, Goicoechea
et al. (2006) showed that the gas sulfur depletion invoked to account for
CS and HCS$^+$ abundances is orders of magnitude lower than in previous
studies of the sulfur chemistry (see also Goicoechea et al. in these
proceedings). In this study, we have made a major step forward as we now
have a non-LTE, non-local (Monte-Carlo) radiative transfer code which
allows us to accurately bridge the gap between the modelled chemical
abundances and the observed spectra. Finally, we observed the
\HthCOp{}~\Jone{} and \DCOp{}~\Jtwo{} and \Jthr{} lines during the winter
and spring 2006. Our aim is to measure the fractional ionization across the
western edge of the Horsehead nebula.  The observed \DCOp{} lines are
surprisingly strong (peak at 3\K{}) inside the dark part of the PDR, just
$\sim$40$''$ away from its illuminated edge.  This opens the interesting
possibility to probe at high resolution the chemical transition from far-UV
photodominated gas to ``dark cloud'' shielded gas in a small field of view.
All those results were unexpected, illustrating the importance of
astrophysical references to constrain chemical models.

\vspace*{-0.20cm}
\section{Conclusion}
\vspace*{-0.25cm}

The PDR on top of the Horsehead nebula has a geometry not only well
understood but also quite simple (almost 1D and viewed edge-on). The
density profile across the PDR is well constrained and there are several
current efforts to constrain the thermal profile. The chemistry of this
source is even richer than expected as it brought numerous surprises.
Finally, its combination of low distance to Earth (400\pc{}), low
illumination ($\chi \sim 60$) and high density ($n \about 10^5\pccm$)
implies that all the interesting physical and chemical processes can be
probed in a field-of-view of less than $50''$ with typical spatial scales
ranging between 1 and $10''$. All those properties make the Horsehead PDR
a source ideally suited 1) for observation with current (\eg{} PdBI) and
future (\eg{} ALMA) radioastronomy interferometers and 2) to serve as
reference to models.



\end{document}